\documentclass[twocolumn]{aastex701}

\usepackage{graphicx}
\usepackage[none]{hyphenat}
\usepackage{amsmath}

\usepackage{multirow}
\usepackage{txfonts}
\usepackage{enumitem}
\usepackage{color}
\usepackage{threeparttable}
\usepackage{algorithmic}
\usepackage{bm}
\usepackage{mathrsfs}
\usepackage{booktabs} 
\usepackage{float}
\usepackage{subfigure}
\DeclareRobustCommand{\VAN}[3]{#2}
\let\VANthebibliography\thebibliography
\def\thebibliography{\DeclareRobustCommand{\VAN}[3]{##3}\VANthebibliography}
\usepackage{color}

\begin{document}

\title{Revealing the Spectral Properties of Galactic Interstellar Medium by Survey Observations}
%%%

\author{Ya-Wen Xiao}
\affiliation{\rm{Department of Physics, Xiangtan University, Xiangtan, Hunan 411105, People's Republic of China}}
\email{ywxwyyx163@163.com}

\author{Jian-Fu Zhang }
\affiliation{\rm{Department of Physics, Xiangtan University, Xiangtan, Hunan 411105, People's Republic of China}}

\affiliation{\rm{National Astronomical Observatories, Chinese Academy of Sciences, 20A Datun Road, Chaoyang District, Beijing 100012, China}}

\email[show]{jfzhang@xtu.edu.cn}

\author{Alex Lazarian}
\affiliation{\rm{Department of Astronomy, University of Wisconsin, Madison, WI, USA}}
\email{jfzhang@xtu.edu.cn}

\begin{abstract}
Based on multi-frequency radio polarization survey datasets, we investigate the spectral characteristics of the Galactic interstellar medium (ISM) using the polarization frequency analysis (PFA) method, referred to as polarization variance. By comparing this novel PFA technique with the traditional power spectrum approach, and by cross-examining data from two distinct surveys, we aim to reinforce the robustness of our findings. Our analysis reveals that the ISM scaling slope in the Galactic disk is steeper than the classic Kolmogorov slope, whereas the ISM scaling slope in the Galactic halo aligns with the Kolmogorov slope. We suggest a distinct turbulence cascade process operating in the Galactic halo compared to the Galactic disk.

\end{abstract}


\section{Introduction} \label{Sect_intro}

The Galactic interstellar medium (ISM), a turbulent magnetized plasma, comprises various phase structures that are widely distributed throughout the Milky Way \citep{Draine2011book}. Magnetohydrodynamic (MHD) turbulence plays a pivotal role in the propagation and acceleration of cosmic rays \citep{Yan2008ApJ, Zhang2021ApJ, LX2023ApJ, Xiao2025AA}, star formation \citep{McKee2007ARA&A, Federrath2012ApJ, Federrath2016ApJ}, magnetic field amplification \citep{XL2016ApJ}, and turbulent magnetic reconnection \citep{LV99, Lazarian2020PhPl, Liang2023ApJ}. Consequently, understanding the properties of the magnetized ISM is critical for deciphering the Milky Way's evolution and various astrophysical processes within it \citep{Beck2013book}. Therefore, accurately extracting ISM information from massive observational datasets—especially the spectral indices—holds significant importance.

Studying the properties of ISM turbulence is a challenging task. Given that the interaction between relativistic electrons and magnetic fields generates synchrotron emission, which provides a theoretical basis for tracing turbulence in the ISM (\citealt{Beck2013book}), one can reveal the fluctuations of the magnetic field in the ISM by synchrotron emission statistics. Previous works have proposed using fluctuation statistics of synchrotron intensity and polarization intensity to reveal magnetic field information in the plane of the sky (\citealt{LP12ApJ, LP2016ApJ}; hereafter LP16). In addition, by combining the Faraday rotation effect, we can trace the magnetic field parallel to the line of sight (LP16, \citealt{Lazarian&Yuen2018ApJ}). However, it is important to note that the presence of the Faraday rotation effect makes the measurement results sensitive to the observational wavelength (\citealt{Lee2016ApJ}), while the measured spectral index is also influenced by the observational spatial resolution (\citealt{Xiao2025ApJ}).

Currently, several methods have been developed to reveal the turbulence spectrum properties of radiation sources. Under the condition of polarization observational data, the spectral characteristics of turbulence can be obtained using the polarization spatial analysis (PSA) and polarization frequency analysis (PFA) techniques (LP16). The latter has been verified by numerical simulations (\citealt{Zhang2016ApJ, Xiao2025AA}). Moreover, the PFA can also reveal the rotation measure spectrum. Additionally, \cite{Guo2024ApJ} confirmed that the PFA technique can be used to measure the strength of magnetic fields.

With the advancement of observational techniques, an increasing number of polarization surveys have provided substantial support for studying the properties of the interstellar medium (ISM) in the Milky Way. Previous polarization observations, such as the northern sky survey conducted with the Dominion Radio Astrophysical Observatory (DRAO) 26-m telescope at 1.4 GHz (\citealt{Wolleben2006A&A}) and the southern sky survey performed with the Villa Elisa 30-m telescope at 1.4 GHz (\citealt{Testori2008A&A}), are single-frequency surveys that provide an all-sky polarization map of the Galaxy. In addition, many polarization surveys—such as the Global Magneto-Ionic Medium Survey (GMIMS; \citealt{Wolleben2009IAUS}), the S-band Polarization All-Sky Survey (S-PASS; \citealt{Carretti2019MNRASS-PASS}), and the Southern Twenty-centimetre All-sky Polarization Survey (STAPS; \citealt{Sun2025A&ASTAPS})—offer continuous multi-frequency observations across the entire sky. This multi-frequency approach helps mitigate the limitations associated with single-frequency observations, thereby improving the reliability of our measurements.

Motivated by the multi-frequency continuous radio polarization sky survey, and in combination with the polarization sky surveys data cubes from STAPS and GMIMS, we investigate the characteristics of the ISM in the Milky Way within the theoretical framework of PFA. One purpose of the current work is to reveal the spectral index of ISM turbulence. Another purpose is to verify the feasibility of the PFA technique for real observational datasets. This paper is organized as follows: Section \ref{Sect_theory} outlines the theoretical foundations and technical methodologies for studying the ISM based on synchrotron radiation; Section \ref{Sect_data} details the survey data cubes utilized in this work; Section \ref{Sect_results} presents the results; and Sections \ref{Sect_discussion} and \ref{Sect_summary} provide discussion and summary, respectively.

\section{Theoretical Considerations}\label{Sect_theory}
\subsection{Synchrotron Polarization}\label{Sect_theory_Syn}
Synchrotron linear polarization, as a pointer characterizing the polarization state of a radiation source, is directly defined by (see LP16 for more details) 
\begin{equation}
P(\bm{X},\lambda^2)=\int_{0}^{L} dz P_i(\bm{X},z)e^{2i\lambda^2\mathrm{RM}(\bm{X},z)}.
\label{eq_PIV}
\end{equation}
It quantifies the relationship of the intrinsic polarized radiation $P_i(\bm{X},z)$ generated at the spatial position $\bm{X}$ and the polarized signal $P(\bm{X},\lambda^2)$ received by the observers under the observational wavelength $\lambda$ when the polarized signal propagates a long distance $L$. During the process, we note that 
\begin{equation}
\mathrm{RM}(\bm{X},z)=0.81 \int_{0}^{z} \rho_e(\bm{X},z')B_\parallel (\bm{X},z')dz' \rm{rad}~\rm{m^{-2}}
\label{eq_RM}
\end{equation} is the Faraday rotation measure (RM) in the ISM, where $\rho_e$ and $B_{\parallel}$ are the density of thermal electrons and the
component of the magnetic field along the line of sight (LOS), respectively. Generally, the synchrotron linear polarization can be calculated by the Stokes parameters $Q$ and $U$ directly, i.e., 
\begin{equation}
P=Q+iU.
\label{eq_PI}
\end{equation} Consequently, the
 synchrotron polarization intensity is $P=\sqrt{Q^2+U^2}$
 (refer \citealt{WaelkensMN2009} for more details).

\subsection{Statistical Methods}\label{Sect_Statistical_Description} 
In this work, we utilize the statistical methods of polarization variance and power spectrum to reveal the underlying properties of turbulence. Under the framework of PFA, from the correlation function of synchrotron polarization intensity $P$, we use a one-point statistical method defining the polarization variance as (LP16):
\begin{equation}
\begin{aligned}
&\langle P(\lambda^2)P^*(\lambda^2) \rangle=\langle P^2(\lambda^2)\rangle=\\
&\int_{0}^{L}dz_1 \int_{0}^{L}dz_2 \times e^{2i\overline{\Phi} \lambda^2 (z_1-z_2)}\langle P_i(z_1)P_i^*(z_2)e^{2i\lambda^2 [\mathrm{RM}(z_1) -\mathrm{RM}(z_2) ]}\rangle,
\label{eq_variance_ori}
\end{aligned}
\end{equation}
where $\overline{\Phi}=\langle \rho_e(z) B_\parallel \rangle$ represents the contribution of the mean RM density.
Combining Equation (\ref{eq_PI}), we can rewrite the polarization variance as:
\begin{equation}
\begin{aligned}
\langle P^2(\lambda^2)  \rangle=&\langle Q^2(\lambda^2)+U^2(\lambda^2)\rangle+i\langle U(\lambda^2)Q(\lambda^2)-Q(\lambda^2)U(\lambda^2)\rangle.
\label{eq_CF_QU}
\end{aligned}
\end{equation}

By comparing the contribution of the mean and fluctuation field on the rotation measure, the predictions of the theory in LP16 classified the relationships of the polarization variance and the square of observational wavelength into the following forms: 
\begin{equation}
\langle P^2(\lambda^2) \rangle \propto \lambda^{-2-2m},
\label{eq_strongFR_mean}
\end{equation}
and 
\begin{equation}
\langle P^2(\lambda^2) \rangle \propto \lambda^{-2},
\label{eq_strongFR_fluctuation}
\end{equation}
corresponding to the case of mean rotation measure dominance and turbulence rotation measure dominance, respectively (see more details in LP16). It is worth noting that $m$ represents the scaling index of the turbulent cascade. Specifically, $m$ equal to $2/3$ means that the cascade satisfies the Kolmogorov-type turbulence cascade.

The power spectrum is a traditional tool to study MHD turbulence, which can reveal the spectral slope, dissipation scale, and many properties of turbulence directly. The power spectrum of any two-dimensional physical quantity $A(\bm{X})$ is
\begin{equation}
PS_{\rm 2D}(\bm{k}) = \frac{1}{(2\pi)^2}\langle A(\bm{X})A(\bm{X+\bm{r}}) \rangle e^{-i \bm{k}\cdot\bm{r}}d\bm{r},
\label{eq_PS_2D}
\end{equation}
where $\bm{r}$ is a separation vector. The ring-integrated 1D spectrum for the 2D variable follows:
\begin{equation}
E(\bm{k}) =\int_{\bm{k-0.5}}^{\bm{k+0.5}} PS_{\rm 2D}(\bm{k})d\bm{k}. 
\label{eq_E_2D}\end{equation}
It is worth emphasizing that $E(\bm{k}) \propto \bm{k}^{-m-2}$, where $m$ is equal to 2/3 for the Kolmogorov spectrum. We have already confirmed that the polarization variance and power spectrum statistics can be cross-validated, i.e., $\langle P^2 \rangle \propto \lambda^{-\delta}$, the 2D power spectrum follows (see \citealt{Xiao2025ApJ} for more details): 
\begin{equation}
E(\bm{k}) = \bm{k}^{-(\delta/2+1)}.
\label{eq_E_k}
\end{equation}

\section{Description of observational datasets}\label{Sect_data}
In our work, we utilize two polarization survey data cubes of $Q$ and $U$ to study the spectral properties of ISM turbulence in the Milky Way based on two survey projects. The first one is from the STAPS (\citealt{Sun2025A&ASTAPS}) by the Parkes 64-m telescopes. This survey covers the sky area with $-89^{\circ} < Dec. < 0^{\circ}$ and for all right ascension, with the frequency ranging from 1.3 to 1.8 GHz at 1 MHz channel. The second one is from the GMIMS (\citealt{Wolleben2019AJGMIMS}) by the CSIRO Parkes 64-m telescope. It almost covers the whole sky with $-180^{\circ} < longitude < 180^{\circ}$ and $-90^{\circ}< latitude < 89.75^{\circ}$ in the Galactic coordinate, with the frequency ranging from 300 to 480 MHz at 0.5 MHz channel bandwidth. The data cubes smoothed are angular resolutions of $20^{\prime}$ (\citealt{Sun2025A&ASTAPS}) and $1^{\circ}.35$ (\citealt{Wolleben2019AJGMIMS}), respectively.

\begin{figure*}
\centering  
\vspace{-0.35cm} 
\subfigtopskip=5pt 
\subfigbottomskip=2pt 
\subfigcapskip=-5pt 
\subfigure{
	\label{level.sub.1}
	\includegraphics[width=1\linewidth]{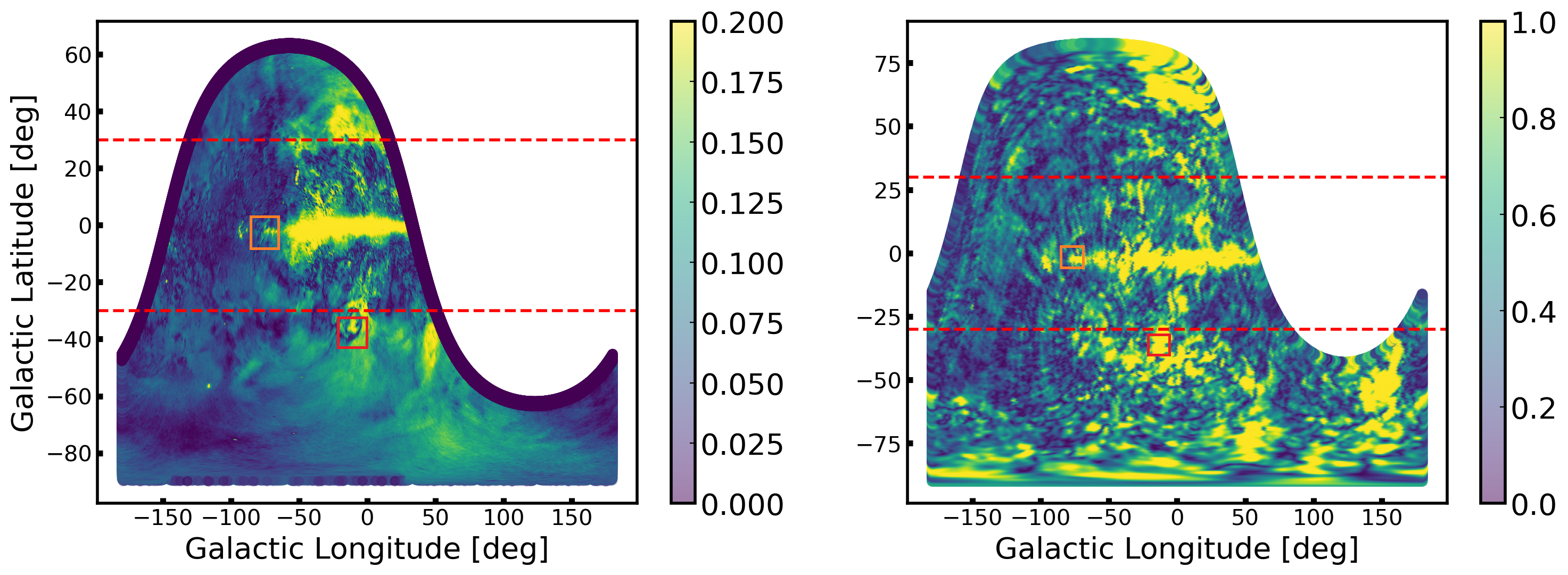}}
\quad 
\caption{Maps of polarization intensities from STAPS (left) and GMIMS (right) corresponding to 1.65 GHz and 0.43 GHz, respectively. The images are in the Galactic coordinates with Plate Carr\'ee projection. The colorbars are in units of $\rm{erg~ s^{-1}~ cm^{-3}~ Hz}$. The horizontal dashed lines represent the dividing line of $\pm 30$ degrees between the Galactic disk and halo. The squared symbols indicate randomly selected subregions located at the Galactic disk (orange) and halo (red).
}
\label{fig: map_PI}
\end{figure*}

With Stokes $Q$ and $U$, we have polarization intensity $P=\sqrt{Q^2+U^2}$ corresponding to STAPS and GMIMS, shown in Figure \ref{fig: map_PI} within the Galactic coordinate, the left and right panels of which are plotted as an example at 1.65 and 0.43 GHz, respectively. Following \cite{Mateu2012MNRAS}, we consider the Galactic disk region in the range of $-30^{\circ} < b < 30^{\circ}$ and the Galactic halo region in the range of $30^{\circ} < b $, as marked out by the horizontal dashed lines. Compared to two maps, we can see that the two surveys reflect similar characteristics of the polarization intensity distributions on the plane of the Milky Way. In particular, the polarization intensity is stronger near the Galactic disk, i.e., $ b\approx 0^{\circ}$, than that of the Galactic halo. It might be due to complex environments such as the presence of a large number of young stars, star clusters, supernova remnants, molecular clouds, and more ionized hydrogen media in the vicinity of the disk. In addition, for high Galactic latitude regions, i.e., $ b\gtrsim80^{\circ}$, given significantly stretching and distortion of the images, we exclude the region of $b\gtrsim80^{\circ} $ in our following statistical analysis. Given that the processes of the energy injection and cascade, as well as the properties of the ISM, have potential differences in the Galactic disk and halo, we perform our analysis for the Galactic disk and halo using the polarization variance technique.

\section{Results}\label{Sect_results}
To study the spectral properties of ISM in the Galactic disk and halo, we first divide the whole observational map into multiple subregions with a size of $15^{\circ} \times 15^{\circ}$ for each frequency channel. We then apply the polarization variance to individual subregions, in comparison with the power spectrum.

\subsection{Single Local Region from Galactic Disk and Halo}

By randomly selecting two subregions located in the Galactic disk and halo (see the squared symbols in Fig. \ref{fig: map_PI}), we explore the feasibility of the polarization variance using both STAPS and GMIMS data cubes, comparing it with the traditional power spectrum. 

\begin{figure*}
\centering  
\vspace{-0.35cm} 
\subfigtopskip=5pt 
\subfigbottomskip=2pt 
\subfigcapskip=-5pt 
\subfigure{
	\label{level.sub.1}
	\includegraphics[width=1\linewidth]{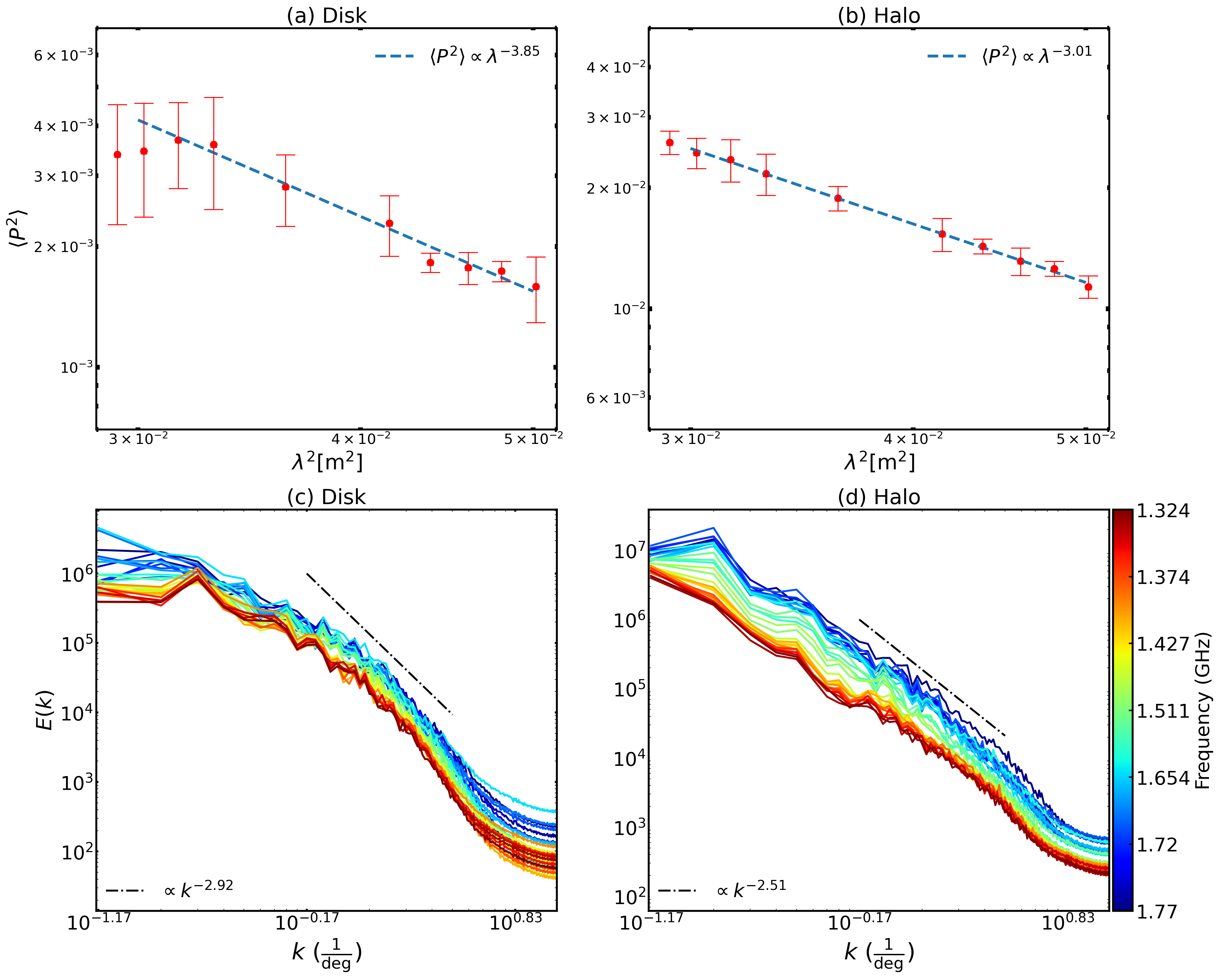}}
\quad 
\caption{Upper panels: polarization variance $\langle P^2 \rangle$ vs. the square of wavelength $\lambda^2$ from two subregions of the Galactic disk and halo. Lower panels: power spectra of polarization intensities corresponding to the two subregions in the upper panels, respectively. The used data cubes are from STAPS. The colorbars show the observational frequency in GHz. 
}
\label{fig: sub_STAPS}
\end{figure*}

Based on the data cubes from STAPS, we show in Fig. \ref{fig: sub_STAPS} the polarization variance $\langle P^2 \rangle$ as a function of wavelength squared $\lambda^2$ (panels (a) and (b)) and the power spectrum of polarization intensity at different observational frequencies (panels (c) and (d)). The error bars plotted in the upper panels represent the standard deviation of the measurement within the corresponding frequency bin. In a subregion of the Galactic disk, we obtain the relationship of $\langle P^2 \rangle \propto \lambda^{-3.85}$ (panel (a)), i.e., $-2-2m=-3.85$, resulting in $m=0.925$, which corresponds to the ISM spectrum of $\propto k^{-0.925}$. From panel (c), we see that the power spectrum follows $E(k) \propto k^{-2.9}$ in the frequency range shown in the colorbar. As the observational frequency decreases, the spectral slope measured by the power spectrum method remains almost unchanged, because depolarization is negligible. Compared with panels (a) and (c), we find that the slope of polarization variance is consistent with that of the power spectrum.

In a subregion of the Galactic halo, we obtain the relation of $\langle P^2 \rangle \propto \lambda^{-3.01}$ (see panel (b)), resulting in $m=0.5$, which is consistent with the power spectrum of $E(k) \propto k^{-2.51}$ in the high frequency range (see panel (d)). As the frequency decreases, we observe that the slope measured by the power spectrum becomes slightly harder, due to the effect of Faraday depolarization. At the same frequency, the radiation of the halo suffers more severe depolarization when compared with panel (c).

\begin{figure*}
\centering  
\vspace{-0.35cm} 
\subfigtopskip=5pt 
\subfigbottomskip=2pt 
\subfigcapskip=-5pt 
\subfigure{
	\label{level.sub.1}
	\includegraphics[width=1\linewidth]{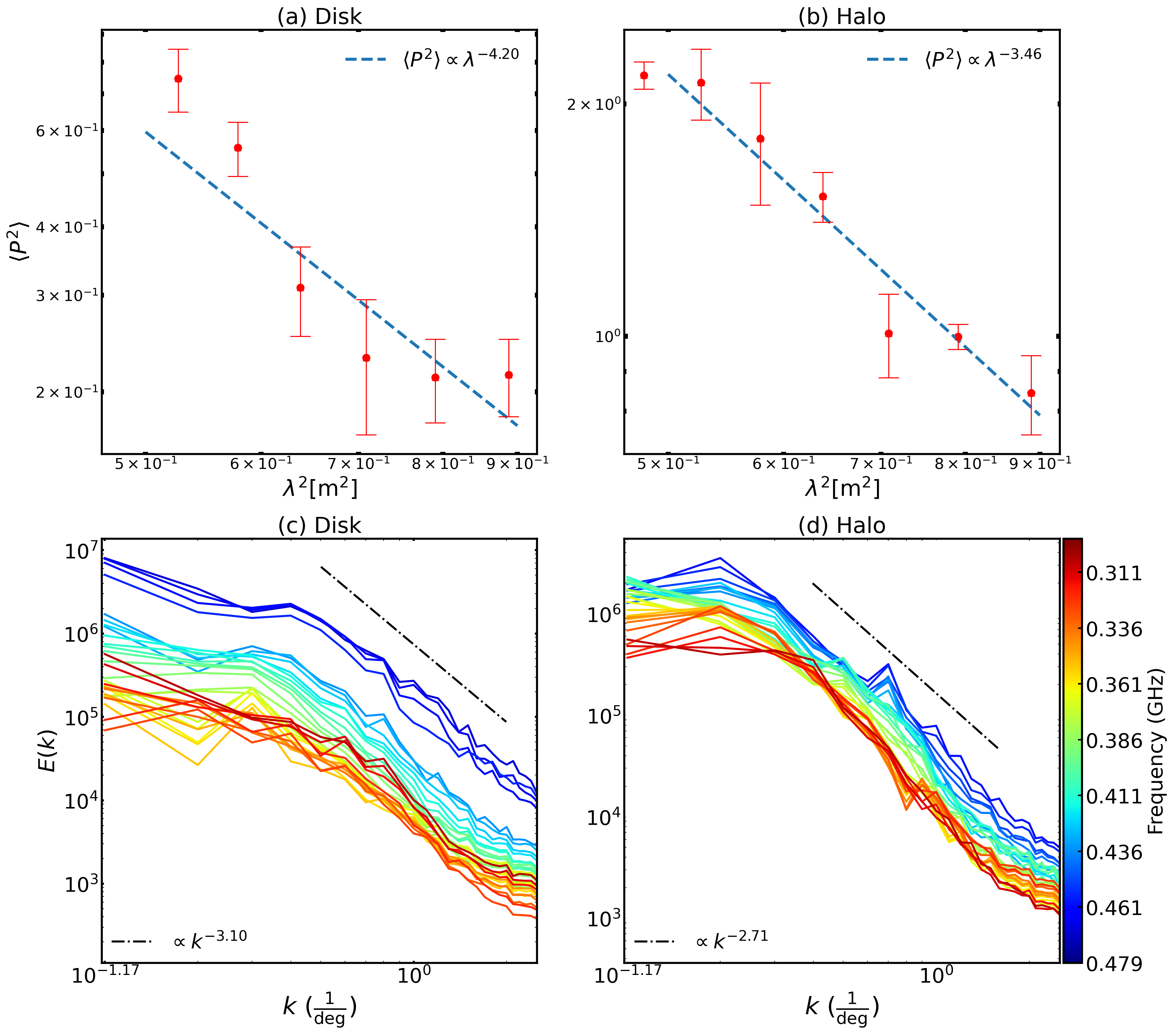}}
\quad 
\caption{Polarization variance $\langle P^2 \rangle$ vs. the square of wavelength $\lambda^2$ (upper panels) and the power spectra of polarization intensity $P$ (lower panels) obtained by GMIMS. The other descriptions are the same as those of Figure \ref{fig: sub_STAPS}.
}
\label{fig: sub_GMIMS}
\end{figure*}

Utilizing the data cubes from GMIMS, Fig. \ref{fig: sub_GMIMS} presents the polarization variance $\langle P^2 \rangle$ vs. the squared wavelength $\lambda^2$ and the power spectrum of the polarization intensities at different observational frequencies. In a subregion of the Galactic disk, we obtain $\langle P^2 \rangle \propto \lambda^{-4.20}$ and $E(k) \propto k^{-3.10}$ at the frequency close to 0.461 GHz in panels (a) and (c), respectively, which means that the scaling index $m$ is close to $1.0$. In a subregion of the Galactic Halo (right panels), we have $\langle P^2 \rangle \propto \lambda^{-3.46}$ and $E(k) \propto k^{-2.71}$ at high observational frequency, resulting in $m \approx 0.7$. 

Compared with the scaling indices of the two subregions, we can see that their slopes within the Galactic disk are steeper than those of the halo, consistent with the results obtained shown in Fig. \ref{fig: sub_STAPS} using the STAPS data cubes. It might be because the energy cascade of ISM in the Galactic disk is influenced by various physical processes such as star formation feedback, supernova explosions, and magnetic reconnection at different scales, causing more energy to accumulate at the large scale.

\subsection{Spectral Statistics of Multiple Local Regions of Galactic Disk and Halo}

\begin{figure*}
\centering  
\vspace{-0.35cm} 
\subfigtopskip=5pt 
\subfigbottomskip=2pt 
\subfigcapskip=-5pt 
\subfigure{
	\label{level.sub.1}
	\includegraphics[width=0.460\linewidth]{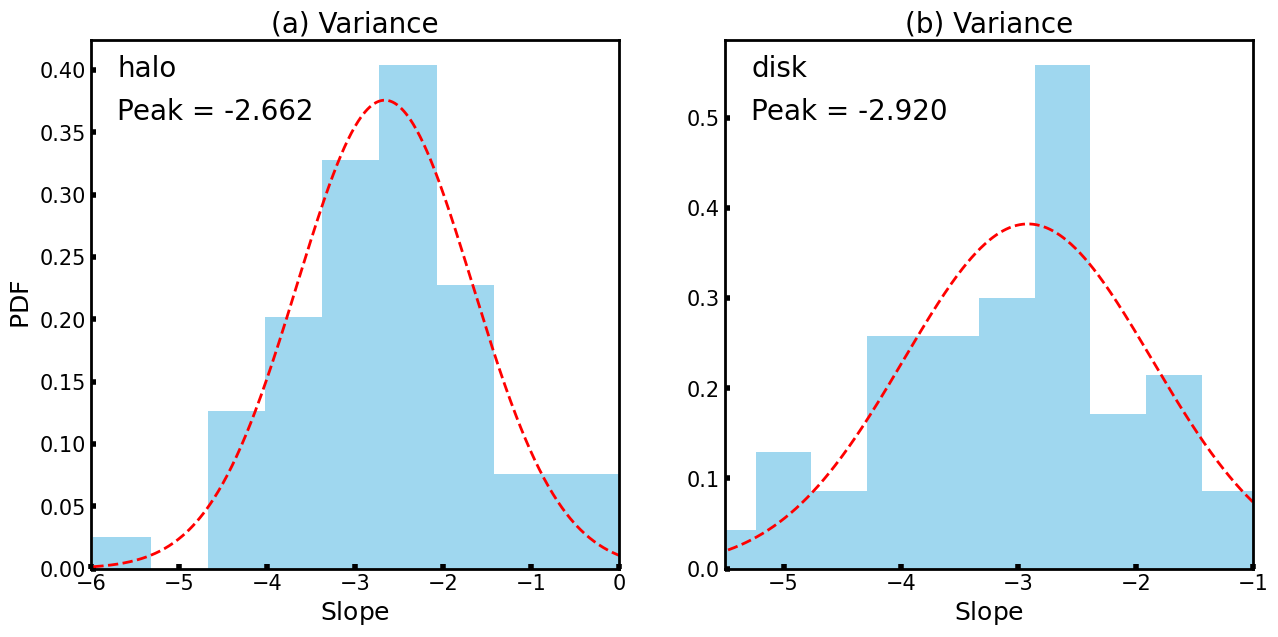}}
\quad 
\subfigure{
	\label{level.sub.2}
	\includegraphics[width=0.450\linewidth]{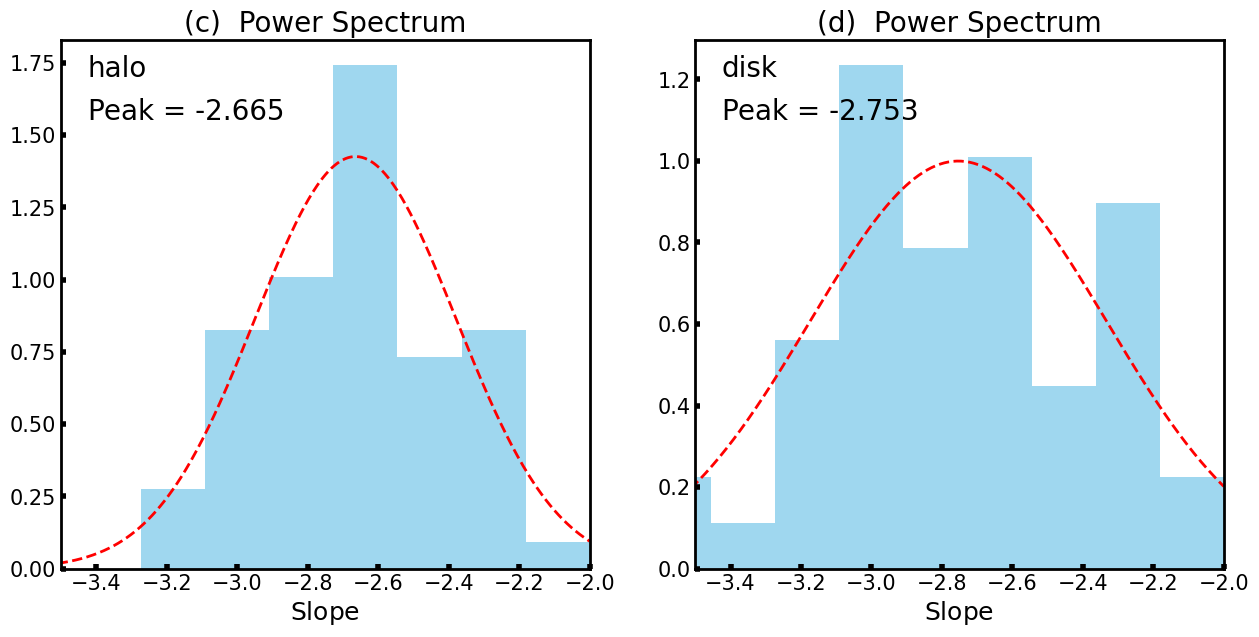}}
\subfigure{
	\label{level.sub.1}
	\includegraphics[width=0.460\linewidth]{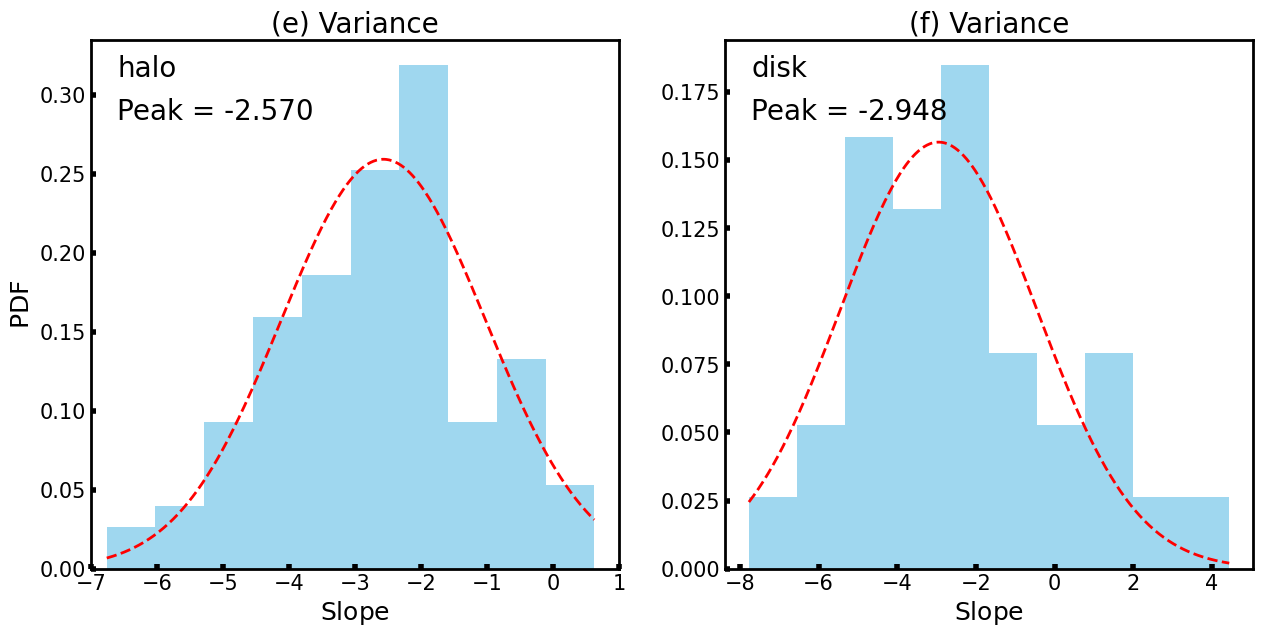}}
\quad 
\subfigure{
	\label{level.sub.2}
	\includegraphics[width=0.450\linewidth]{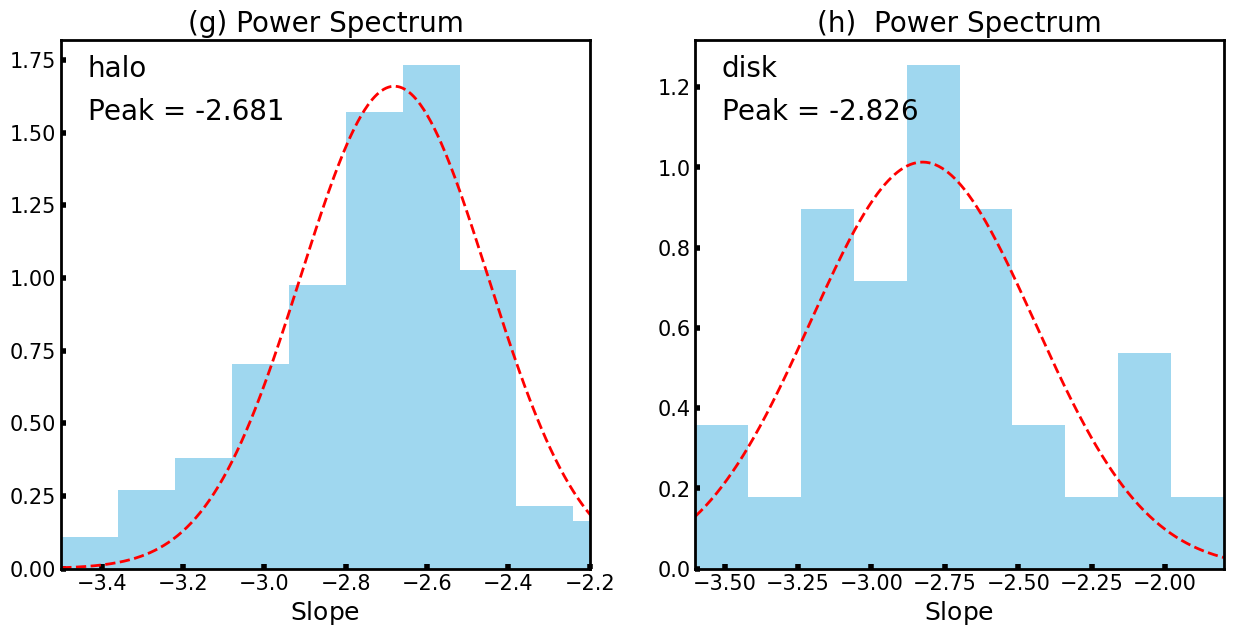}}
\caption{Probability density function (PDF) of spectral indices of magnetized ISM corresponding to the Galactic disk and halo, using the polarization variance and power spectrum of polarization intensities, from STAPS (upper panels) and GMIMS (lower panels) data cubes.}
\label{fig:halo_disk}
\end{figure*}

Given the possibility of uncertain results for a single randomly selected local region, we explore the statistical properties of multiple local regions in this section for more reliable results. Fig. \ref{fig:halo_disk} shows the distribution of spectral indices of magnetized ISM, with the Gaussian fitting, derived from polarization variance, in comparison with the power spectrum. Using the data cube from STAPS, we obtain that the peaks of distribution are 2.662 (panel (a)) and 2.920 (panel (b)), corresponding to the Galactic halo and disk, respectively. As seen in panels (c) and (d), the peaks of distribution from the power spectrum are 2.665 and 2.753, corresponding to the Galactic halo and disk, respectively. Moreover, using the data cube from GMIMS, we obtained peak values of 2.570 and 2.948 by the polarization variance (see panels (e) and (f)), as well as 2.681 and 2.826 by the power spectrum (panels (g) and (h)).

The spectral index distributions demonstrate that the Galactic halo’s magnetized ISM turbulence is consistent with the 2D Kolmogorov scaling, while the disk’s turbulence deviates significantly, exhibiting steeper spectral indices. These differences reflect the distinct physical environments of the two regions: the halo is a more quiescent, magnetically dominated medium, while the disk is a dynamically active region with strong shocks, shear, and compressible effects that modify the turbulent cascade. The consistency between different methods and datasets further validates the use of the polarization variance technique to probe ISM turbulence.

\section{Discussion}\label{Sect_discussion}

The comparison between polarization variance and power spectrum methods, as well as between STAPS and GMIMS datasets, highlights the robustness of the measured results. For the halo, the two methods yield nearly identical peak values (e.g., 2.662 vs. 2.665 for STAPS), confirming the reliability of the spectral slope measurements. For the disk, the methods show slightly larger differences (e.g., 2.920 vs. 2.753 for STAPS), likely due to the disk’s more complex ISM and strong Faraday depolarization, which introduces different systematic biases in the power spectrum method. Both STAPS and GMIMS show the same qualitative trend: halo indices near Kolmogorov-type turbulence of 8/3, and disk indices steeper than 8/3. Our cross-validation strengthens the potential physics that the halo and disk have distinct turbulent properties.

For the Galactic halo, measurements from both the STAPS and GMIMS datasets show remarkable alignment with the theoretical 2D Kolmogorov index. This near-universal agreement suggests that the Galactic halo’s ISM turbulence is dominated by \cite{GS95}'s incompressible cascade modes, with minimal disruption from strong magnetic fields or compressible shocks.

In contrast, the Galactic disk exhibits a consistent departure from the 2D Kolmogorov prediction, with all measured spectral indices being steeper than 8/3. This steepening implies that the disk’s ISM turbulence deviates from the canonical Kolmogorov cascade. Some key physics for this deviation may include--(1) stronger magnetic fields and higher plasma $\beta$: the disk’s higher magnetic field strength and thermal pressure can break the self-similarity of the energy cascade, leading to steeper spectra; (2) compressible effects: supernova remnants, shocks, and density inhomogeneities in the disk introduce compressible turbulence, which modifies the inertial range scaling; (3) complex velocity fields: the disk’s rotational shear and large-scale flows can disrupt the isotropic, homogeneous assumptions of Kolmogorov theory.

From a technical perspective, to reflect the relationship between polarization variance and the squared wavelength, we divide the frequency from 1.3 to 1.8 GHz into 10 logarithmic bins and from 300 to 480 MHz into 6 logarithmic bins, corresponding to STAPS and GMIMS datasets, respectively. We have tested that the change in the selected bin number did not distort the statistical results. In addition, in the process of obtaining the power law of the power spectrum, for each subregion, we used a consistent upper and lower boundary of the inertial range, which corresponds to the injection and dissipation scales of turbulence, respectively, to ensure that the execution of the method is universal. However, due to the uncertainty of the injection and dissipation scales, the power spectrum method exposes its own shortcomings. Interestingly, the polarization variance method gets rid of dependence on injection and dissipation scales.

In the framework of traditional Faraday rotation synthesis, \cite{Burn1966MNRAS} predicted a strong depolarization relation of $\Pi \propto e^{-\lambda^4}$ between the linear polarization degree and the wavelength. We noted that \cite{Burn1966MNRAS} did not discuss turbulence cascades by considering the frozen magnetic fields. After considering the turbulence cascade process, \cite{Tribble1991MNRAS} proposed a shallower relation of $\Pi \propto \lambda^{-4/\zeta}$, where $\zeta$ is the spectral index of the turbulence cascade. In contrast, our current work focuses on the power-law relationship between the polarization intensity variance and the wavelength. Exploring the underlying relationship between the polarization degree and the wavelength will be carried out in the framework of the modern understanding of MHD turbulence in future work.

Based on \cite{Tribble1991MNRAS}’s depolarization relation, \cite{Mao2017ApJ} established the connection between the turbulence spectral index $\zeta$ and the X-ray to optical power-law spectral index $\beta$ by $\zeta=\beta+1$ in the polarization of GRB 121024A. As defined in \cite{Mao2017ApJ}, $\zeta$ is related to the scaling index of the two-point multi-order structure function and can characterize the intermittency of magnetic turbulence. In fact, the scaling slope of multi-order structure function can reveal well the feature of small-scale turbulence, as pointed out by \cite{Wang2024A&A}. Given the spectral and spatial resolution of observational cubic data, we cannot obtain turbulence information in the range of kinetic scales, which has been archived in the solar wind turbulence (\citealt{Sahraoui2009PhRvL}). Therefore, our current work focused on a large-scale strong turbulence regime in the inertial range. Our finding is important for understanding the large-scale turbulence cascade from the Galactic disk and halo.

The synchrotron polarization intensity statistics explored in this paper are to obtain the cascading power law of the magnetic field. To fully reveal the spectral properties of Galactic magnetized ISM, using spectroscopic datasets is a good way to obtain the spectral properties of the ISM velocity. In this regard, there are relatively mature statistical methods such as the techniques of velocity channel analysis and velocity coordinate spectrum (\citealt{LP2000ApJ, LP2006ApJ}).

We confirmed that polarization variance serves as an effective tool for measuring the magnetic field spectral index of the Galactic ISM. Additionally, the structure function can also be employed for this purpose. By statistically analyzing the structure function of RM, \cite{Xuzhang2016ApJ} found that the RM exhibits a shallower spectrum in the Galactic disk, whereas the measured RM spectrum in the halo aligns with the Kolmogorov spectrum. In light of these findings, we speculate that a shallow density spectrum in the disk region may contribute to the observed shallowness of the RM spectrum (see Fig. 9 of \citealt{Seta2023MNRAS} for numerical simulations).

\section{Summary}\label{Sect_summary}
In this paper, we focused on the application of the polarization variance to the slope measurements of magnetized ISM within the Galactic disk and halo. To verify the reliability of the polarization variance, we conduct a cross-validation between measurement methods (together with the power spectrum) and observational datasets (STAPS and GMIMS). Our main findings are summarized as follows:

\begin{enumerate} %[itemindent=2.0em]
\item We have already demonstrated that polarization variance can be successfully applied to real observational datasets, which paves the way for applying this technique to the massive SKA dataset to understand the evolution of the cosmic magnetic field.

\item When revealing the spectral characteristics of ISM turbulence, the polarization variance depends on polarization multi-frequency analysis, which is better suited to the current and future polarization observational dataset patterns.

\item The advantage of the polarization variance over power spectrum measurements is that it is not affected by Faraday depolarization, and it does not need to predetermine the injection and dissipation scales, which are difficult to determine, when obtaining the slope in the inertial cascade range.

\item In the case of the Galactic disk, due to the complex Galactic environment, the spectrum of ISM turbulence is steeper than the Kolmogorov spectrum. In the case of the Galactic halo, the spectrum of the ISM turbulence follows the Kolmogorov turbulence.

\end{enumerate} 

The datasets used for Stokes $Q$ and $U$ from STAPS and GMIMS can be accessed from the Canadian Advanced Network for Astronomical Research via \citet{https://doi.org/10.11570/25.0004} and \citet{https://doi.org/10.11570/18.0007}, respectively.

\begin{acknowledgments}
We thank the anonymous referee for valuable comments that significantly improved the quality of the paper. J.F.Z. acknowledges the support from National SKA Program of China (2025SKA0140100), and the National Natural Science Foundation of China (grant No. 12473046). A.L. acknowledges the support of 1091 NSF grants AST 2307840.
\end{acknowledgments}
\vspace{5mm}
\bibliography{sample701}{}
\bibliographystyle{aasjournal}

\end{document}